\documentclass[12pt]{article}
\usepackage{graphicx, epsfig, color}
\usepackage{amssymb,amsmath}

\def\eslt{E_T^{\rm miss}}

\def\to{\rightarrow}

\def\bi{\begin{itemize}}
\def\ei{\end{itemize}}

\def\th{\tilde h}

\def\tg{\tilde g}

\def\tw{\widetilde W}
\def\tz{\widetilde Z}

\def\be{\begin{equation}}  
\def\ee{\end{equation}}  
\def\bea{\begin{eqnarray}}  
\def\eea{\end{eqnarray}}  
\def\beas{\begin{eqnarray*}}  
\def\eeas{\end{eqnarray*}}

\newcommand\prd[3]{{\it Phys.\ Rev.\ }{\bf D#1} (#2) #3}

\newcommand\prl[3]{{\it Phys.\ Rev.\ Lett.\ }{\bf #1} (#2) #3}
\newcommand\plb[3]{{\it Phys.\ Lett.\ }{\bf B #1} (#2) #3}
\newcommand\jhep[3]{{\it J. High Energy Phys.\ }{\bf #1} (#2) #3}

\newcommand\epjc[3]{{\it Eur.\ Phys.\ J. }{\bf C #1} (#2) #3}

\begin{document}
\begin{titlepage}
\begin{flushright}
UH-511-1246-15
\end{flushright}

\vspace{0.5cm}
\begin{center}
{\Large \bf  Same-sign Higgsino Production at the CERN LHC:\\ How Not to
  Hunt for Natural Supersymmetry} 
\\ 
\vspace{1.2cm} \renewcommand{\thefootnote}{\fnsymbol{footnote}}
{\large 
Patrick Stengel$^1$\footnote[1]{Email: pstengel@hawaii.edu} and Xerxes Tata$^1$\footnote[2]{Email: tata@phys.hawaii.edu } 
}\\ 
\vspace{0.5cm} \renewcommand{\thefootnote}{\arabic{footnote}}
  
{\it 
$^1$Dept. of Physics and Astronomy,
University of Hawaii, Honolulu, HI 96822, USA \\
}
\end{center}

\vspace{0.5cm}
\begin{abstract}
\noindent 
\vspace*{0.3cm}

We examine the prospects for detecting light charged higgsinos that are
expected to be a necessary feature of natural SUSY models via $pp\to
\tw_1^\pm\tw_1^\pm jj+X$ processes arising dominantly from $W^\pm W^\pm$
fusion at LHC13. The signal will be a pair of same-sign leptons ($e$ or
$\mu$) in events with two relatively forward, hemispherically-separated
jets with a large rapidity gap. We find that even though the higgsinos
have a full-strength $SU(2)$ gauge couplings to $W$-bosons, the LHC13
cross section for the production of same sign higgsino pairs is smaller
than 0.02~fb over most of the interesting range of natural SUSY
parameters, even before leptonic branching fractions of the chargino are
included.  This cross section is strongly suppressed because the two
neutral Majorana higgsinos can be combined into a single Dirac
neutralino if the bino and the winos are much heavier than the
higgsinos, as is the case in natural SUSY models: in this limit,
higgsino couplings to $W$-bosons exhibit an emergent (approximate)
$U(1)_{\rm ino}$ global symmetry that suppresses same sign higgsino
production by vector boson fusion.  We conclude that this channel is not
a viable way to search for natural SUSY even at the high luminosity
upgrade of the Large Hadron Collider.

\end{abstract}

\end{titlepage}

Despite the absence of any signals from direct production of
superpartners in experiments at LHC8 \cite{atlas_susy,cms_susy}, weak
scale supersymmetry (SUSY) remains the most promising extension of the
Standard Model (SM). That interest in space-time supersymmetry continues
more than four decades after its discovery \cite{wessz} is largely
driven by the fact that softly-broken SUSY stabilizes \cite{hier} the
Higgs sector of the SM from run-away radiative corrections that arise
\cite{susswein} when the SM is embedded into a framework with a
hierarchically different mass scale such as Grand Unification.  While it
is clear that supersymmetry elegantly resolves the big hierarchy problem
if superpartners are at the weak scale, several authors \cite{ftune}
have expressed concern that LHC8 bounds on top squarks may be already
indicative of fine-tuning at the percent level unless the SUSY spectrum
happens to be compressed so that signals from $t$-squarks would have
evaded experimental searches \cite{lhcstop}. It has, however, been
pointed out \cite{rnsltr,rns,howieover,am} that these authors have
implicitly ignored the possibility that various SUSY parameters might be
correlated, and that it may be possible to find models with relatively
modest fine-tuning, but with top squarks in the TeV range, well beyond
the reach of LHC experiments. We note instead that a small value of the
superpotential $\mu$ parameter (assuming that it arises from a different
origin than the SUSY breaking parameters) which directly enters the
Higgs potential, and through it in the well-known expression for $M_Z^2$
(see, {\it e.g.} Eq.~(4) of the second paper of Ref.~\cite{am}),
provides a necessary condition for low fine-tuning; see also
Ref.\cite{CCN}. We thus advocate light higgsinos in the 100-300~GeV
range as the most robust feature of natural SUSY models without a
proliferation of new particles beyond the MSSM.\footnote{Very recently
models with additional chiral superfields in the adjoint representation
\cite{nelmar}, or with additional superfields needed to complete
representations of a large global symmetry group so that the Higgs is a
pseudo-Goldstone boson \cite{luty}, where the higgsino mass is
independent of the $\mu$ parameter that enters the Higgs boson sector
have been constructed. In these relatively complex scenarios, higgsinos
may be heavy consistent with low electroweak fine-tuning.}

Electroweak higgsino pair production via $pp \to \tz_i\tz_j$,
$\tw_1^+\tw_1^- \ {\rm and} \ \tw_1^\pm\tz_i+X$ processes has a cross
section of several hundred to over a thousand fb at LHC13 for higgsinos
in the natural SUSY mass range of 100-300~GeV \cite{rnslhc}. However,
since electroweak gaugino mass parameters can easily be in the TeV range
without endangering naturalness, the mass gaps $m_{\tw_1}-m_{\tz_1}$ and
$m_{\tz_2}-m_{\tz_1}$ are typically just 10-30~GeV for electroweak
fine-tuning no worse than 3\% \cite{ilc}. As a result, the visible
energy and $\eslt$ in these higgsino production reactions is small, and
higgsino production is obscured by SM backgrounds.\footnote{It is
possible that bino and wino mass parameters have a magnitude comparable
to $|\mu|$. In this fortuitous case the resulting substantial mixing
between the gauginos and higgsinos typically splits the various states
and leads to the possibility of several observable signals at the LHC
\cite{nugm}.} This led several groups to examine the possibility of
detecting the higgsino signal via higgsino pair production in
association with a high $p_T$ jet or photon from 
QCD/QED radiation. Careful
studies have shown that while the signal occurs with an observable rate
after hard cuts on $\eslt$ and the hard jet/photon, the signal to
background ratio is at the 1-2\% level \cite{mono}. It seems difficult
to imagine that the systematic error on the QCD backgrounds will be
smaller than this level, strongly indicating that the signal will be
difficult to detect in the absence of characteristic kinematic features
in these events. In Ref.\cite{carp}, it has been suggested that higgsino
production with a radiated $Z$ boson decaying to a lepton pair provides
a better reach. Taking the results of this study at face value, there is
no observable signal for an integrated luminosity of 300~fb$^{-1}$ at
LHC13, and a $5\sigma$ signal for higgsino masses up to 125-130~GeV for
an integrated luminosity of 3000~fb$^{-1}$. Following earlier work in
Ref.\cite{ghww}, Han {\it et al.} \cite{kribs} suggested that it may be
possible to reduce SM backgrounds by requiring soft dileptons in
higgsino pair production in association with a hard monojet. A
subsequent detailed study \cite{dilep} showed that by requiring low
invariant mass, opposite sign, same flavour dileptons in hard mono-jet
events, it is possible to extract the higgsino signal above SM
backgrounds at LHC14 for $|\mu| < 170$ (200)~GeV, assuming an integrated
luminosity of 300 (1000)~fb$^{-1}$. Though this covers the range of
$\mu$ most favoured by naturalness considerations, the strategy does not
lead to observability over the entire range of $\mu$ allowed by 3\%
electroweak fine-tuning. While signals from the production of higgsinos
will be readily detectable at an electron-positron collider with
$\sqrt{s} > 2m_{\rm higgsino}$ and availability of electron beam
polarization \cite{ilc}, it is clear that alternative strategies to
search for these at the LHC are worthy of examination.

The ATLAS \cite{atlasWW} and CMS \cite{cmsWW} collaborations 
have recently reported observation of same-sign (SS) $W$-pair
production via $W^\pm W^\pm$ scattering at LHC8. This,  together
with the fact that charged higgsinos of natural SUSY must have masses
not hierarchically larger than $M_W$, motivated us to examine SS
charged higgsino pair production via $pp \to \tw_1^\pm\tw_1^\pm jj+X$ at
LHC13 in this paper. The signal is a pair of same-sign dileptons
(from the leptonic decays of the charginos) together with forward
hemispherically separated jets with a large rapidity separation and
$\eslt$.  Unlike the leptons from $W$ decay which would typically have
$p_T \sim M_W/2$, the leptons from $\tw_1$ decays are expected to be
soft because $m_{\tw_1}-m_{\tz_1}$ is expected to be just 10-25~GeV in
natural SUSY models.  It would then be of interest to examine if it is
possible to search for a signal in events triggered by high $E_T$
forward jets with a large rapidity separation, with a pair of soft,
acollinear dileptons and modest $\eslt$.

\begin{figure}[tbh]{\begin{center}
\includegraphics[width=10cm,clip]{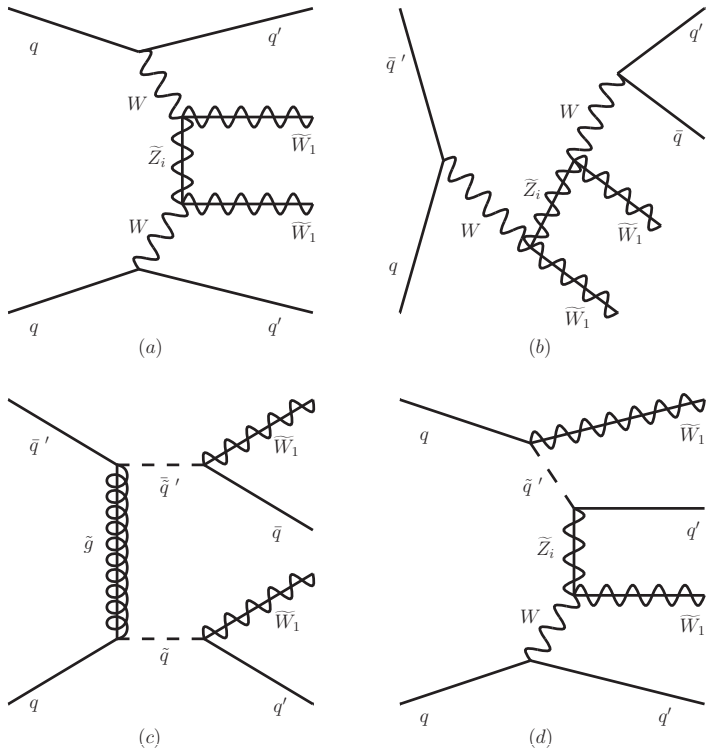}
\caption{Representative Feynman diagrams for the underlying parton 
processes that contribute to the production of same-sign charginos
in $pp$ collisions at LHC13. As explained in the text, we that 
contributions from the processes {\it b-d} to be strongly suppressed.}
\label{fig:SSw1w1}
\end{center}}
\end{figure}

SS chargino pair production occurs via the SUSY analogues of the Feynman
diagrams that lead to inclusive $W^\pm W^\pm jj$ production at the LHC.
Representative examples are illustrated in Fig.~\ref{fig:SSw1w1}.
Fig.~\ref{fig:SSw1w1}{\it a} shows the classic vector boson fusion (VBF)
diagram for SS chargino production. Fig.~\ref{fig:SSw1w1}{\it b} shows
the $s$-channel $W^*$ diagram. This is expected to be suppressed once we
require the hard, hemispherically separated jets (see below)
characteristic of VBF events. Fig.~\ref{fig:SSw1w1}{\it c} shows a
mixed QCD-electroweak diagram which is strongly suppressed in natural
SUSY because squarks and gluinos are expected to be heavy. Finally,
Fig.~\ref{fig:SSw1w1}{\it d} illustrates a purely electroweak diagram
that is also suppressed if squarks are heavy.

Superpartner production via VBF processes was first examined almost a
decade ago in Ref.\cite{tilman}. It has more recently received attention
in Ref.\cite{ghww}, and most extensively in a series of papers by the
Texas A and M group \cite{aggies}. Since $\tw_1^\pm\tw_1^\mp$ and
$\tz_i\tz_j$ production in association with high transverse momentum
jets also occurs via conventional quark-antiquark initiated processes,
we confine our examination to production of just same-sign chargino
pairs in this paper.

Since our focus is the SS higgsino signal in natural SUSY, for
definiteness, we adopt the Radiative Natural SUSY (RNS) model line
developed in Ref.\cite{rnslhc} for our calculations.\footnote{Assuming
that the electroweak gauginos are not fortuitously light, we expect that
the same sign higgsino production cross section is largely determined by
the magnitude of $\mu$, and largely independent of the details of the
model.}  The RNS framework zeroes in on the portion of parameter space
of the non-universal Higgs mass model with two additional parameters
(NUHM2 model) beyond those of the well-studied mSUGRA/CMSSM framework
\cite{nuhm2}. The NUHM2 model which is
defined by the parameter set,
$$m_0,m_{1/2},A_0, \tan\beta,\mu, m_A,$$
allows for spectra with modest electroweak fine-tuning\footnote{As
  explained at length in Ref.\cite{am}, we regard low $\Delta_{\rm EW}$
  as a necessary (but not sufficient) condition for low fine-tuning in
  any SUSY model.}  ($\Delta_{\rm EW}^{-1} \ge 3$\%), consistent with
  the observed Higgs mass \cite{rns}, if parameters are appropriately
  correlated. Specifically, we first fix, \be
m_0=5~{\rm TeV},  A_0=-1.6m_0,
\tan\beta=15, \mu=150~{\rm GeV}, m_A=1~{\rm TeV},
\label{eq:mline}
\ee 
and examine the signal versus $m_{1/2}$ which we allow to vary up to
2~TeV.\footnote{For this model line, the electroweak fine-tuning
parameter $\Delta_{\rm EW} < 30 \ (80)$ if $m_{1/2} < 1.2 \ (2.0)$~TeV.} The
higgsino masses, $m_{\tw_1}$ and $m_{\tz_{1,2}}$, are essentially fixed
by $|\mu|$ (as long as $m_{1/2}$ is chosen large enough so that the weak
scale gaugino parameters $M_{1,2} \gg |\mu|$).  Given heavy sfermions expected in natural SUSY, our results are
insensitive to the specific values of $m_0$, $\tan\beta$ and $m_A$.

We use the subroutines in ISAJET v7.83 \cite{isajet} to compute the
sparticle spectrum (illustrated in Fig.~2 of Ref.\cite{rnslhc}) and
mixing parameters. We then use Madgraph5.2.2.1 \cite{madgraph} to
evaluate the squared matrix element for the $2\to 4$ parton subprocess, and feed the result into Pythia v6.426 \cite{pythia} where we
convolute the partonic cross section with CTEQ6L1 distribution functions
\cite{cteq}, implement the $p_T$-ordered shower using the $k_T$
clustering scheme to form jets, and obtain the cross section at LHC13.
We use PGS4 \cite{pgs} for a toy simulation of the detector in the LHC
configuration, and define jets to be hadronic clusters with $E_T(j)>
30$~GeV within $|\eta_j| < 5$.

\begin{figure}[tbh]
\begin{center}
\includegraphics[width=12cm,clip]{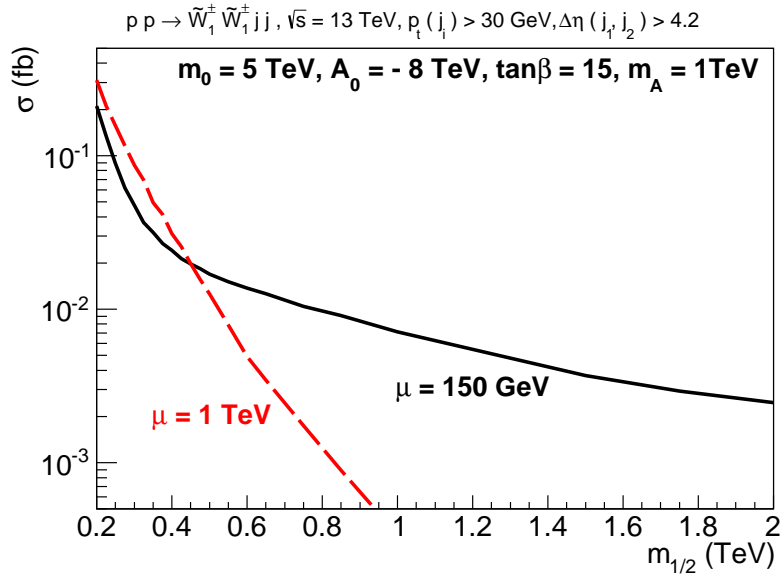}
\caption{The cross section for SS chargino production via $pp\to
  \tw_1^\pm\tw_1^\pm jj+X$ at
a centre-of-mass energy $\sqrt{s} =13$~TeV versus $m_{1/2}$ for the RNS
  model line (\ref{eq:mline}) of the text (solid, black curve) and for
  the same model-line but with $\mu=1$~TeV (dashed, red curve).}
\label{fig:SSw1w1vsmhalf}
\end{center}
\end{figure}

The solid (black) curve in Fig.~\ref{fig:SSw1w1vsmhalf} shows the result
of our computation of $\sigma(pp \to \tw_1^\pm\tw_1^\pm jj+X)$ versus
$m_{1/2}$ for the natural SUSY model-line in (\ref{eq:mline}). 
In this figure we have required:
\bi
\item At least two jets with $E_T(j) > 30$~GeV, and
\item  the rapidity separation between the two highest $E_T$
  jets, $\Delta\eta(j_1,j_2)>4.2$.
\ei
We have checked that requiring the jets to be in opposite hemispheres so
  that $\eta(j_1)\cdot\eta(j_2) < 0$ does not affect the result. 
The
range with $m_{1/2} \lesssim 475$~GeV is, of course, excluded by the lower bound $m_{\tg} >
1.3$~TeV \cite{atlas_susy,cms_susy}. For small values of $m_{1/2}$, the
chargino is a mixed gaugino-higgsino state, but becomes almost a pure
higgsino with a mass $m_{\tw_1} \simeq \mu = 150$~GeV once $m_{1/2}$
exceeds 500-600~GeV.  The surprising feature is that the cross section
drops off to below 0.02~fb for $m_{1/2} \ge 500$~GeV and continues to
fall with increasing $m_{1/2}$ {\em even though the chargino mass,
remains fixed close to $\mu$ across most of the range of $m_{1/2}$ where
$\tw_1$ is dominantly higgsino-like.} For comparison, we also show by
the dashed (red) line the cross section for the same model line but
with $\mu=1$~TeV for which the lighter chargino is mostly
wino-like. This cross section also drops off with increasing $m_{1/2}$
but this fall-off is clearly because the wino mass increases with
$m_{1/2}$ according to $M_2\simeq 0.8m_{1/2}$.

\begin{figure}[tbh]{\begin{center}
\includegraphics[width=15cm,clip]{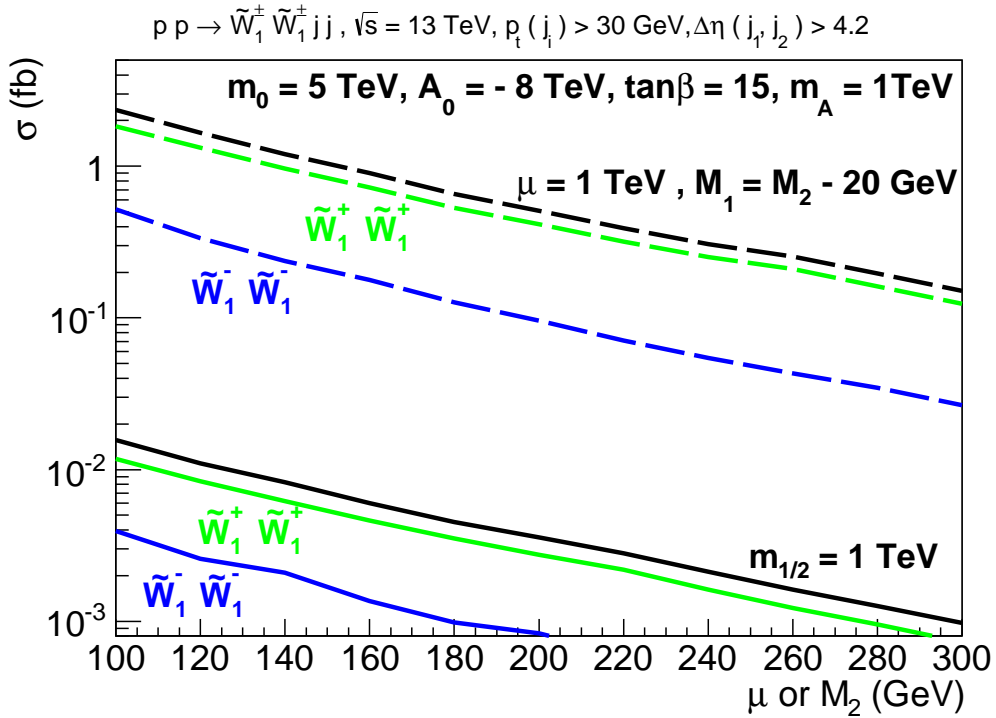}
\caption{The cross section for SS chargino production via $pp\to
  \tw_1^\pm\tw_1^\pm jj+X$ at a centre-of-mass energy $\sqrt{s} =13$~TeV
  versus $\mu$ for the RNS model line (\ref{eq:mline}) of the text with
  $m_{1/2}=1$~TeV (solid curves), and versus $M_2$ with $M_1=M_2-20$~GeV
  and $\mu=1$~TeV (dashed curves), with other weak scale parameters at
  the same values used for the solid curves. The uppermost of each set
  of curves shows the total SS chargino cross section, while the other
  two curves in each set denote the component cross sections. Plotted
  this way, the scale on the horizontal axis is, to a good
  approximation, the mass of the lighter chargino for both sets of
  curves in the figure.}
\label{fig:SSw1w1vsmu}
\end{center}}
\end{figure}

To illustrate the dependence of the SS chargino production cross section
in natural SUSY on the chargino mass, we show $\sigma(pp\to
\tw_1^\pm\tw_1^\pm jj+X)$ versus $\mu$ in Fig.~\ref{fig:SSw1w1vsmu} for
the same model-line but with $m_{1/2}=1$~TeV (uppermost solid
curve).  The lighter chargino is dominantly a higgsino along this curve.  Also shown are the component cross sections for $\tw_1^+\tw_1^+jj$
events and $\tw_1^-\tw_1^-jj$ events at the LHC. The production of
positive chargino pairs exceeds that of negative chargino pairs because
there are more up type than down type quarks in a proton.  We also see
that the total same sign chargino production cross section lies below
0.01~fb over essentially the entire range of the plot. The dashed
lines in the figure illustrate the cross sections for the same reaction,
but in a model without gaugino mass unification where the lighter
chargino is essentially a pure wino.  We show this cross section versus
the wino mass parameter $M_2$, with $M_1=M_2-20$~GeV and $\mu=1$~TeV,
with values of other {\em weak scale parameters} the same as those for
the solid curves.
We see that the cross section for SS higgsino pair production is about
two orders of magnitude smaller than that for SS wino pair production
for the same mass of the particle.\footnote{We have checked that our
results for same sign chargino production are compatible with results in
Ref.\cite{ghww} but not with those in Ref.\cite{tilman}. We also obtain
a same sign chargino cross section that is a factor of about 20 smaller
than the uppermost curve in Fig.~2 of the second paper of Ref.
\cite{aggies}. We have contacted these authors who have since confirmed
that they agree with our result. We thank B.~Dutta, T.~Ghosh,
A.~Gurrola, T.~Kamon and especially T.~Plehn and S.~Wu for extensive
communications and discussion concerning the discrepancy.} Indeed the
latter occurs at an observable rate even after factoring in the
branching fraction of 0.22 for the leptonic decay of each chargino, for
an integrated luminosity of 300 (3000)~fb$^{-1}$ expected to be
accumulated at LHC Run 2 (the high luminosity upgrade of the LHC). In
contrast, the cross section for SS higgsino production appears to be
below the level of observability at LHC13:\footnote{This situation may
be different at a 100~TeV $pp$ collider (S$pp$C) being envisioned for
the distant future \cite{sppc}. Assuming that the $\tw_1^\pm\tw_1^\pm
jj$ cross section scales by the same factor as the corresponding
neutralino cross section from VBF \cite{liantao} between the LHC and the
S$pp$C, we may expect $\sim 2$ SS dilepton + $jj$ events per ab$^{-1}$
for a charged higgsino mass of 200~GeV and $m_{1/2}=1$~TeV. For an
integrated luminosity of 30~ab$^{-1}$ \cite{sppc}, this corresponds to
$\sim 60$ events before any acceptance, trigger and analysis cuts, or
efficiency corrections. We make no representation about the
observability of the SS chargino signal via the dilepton plus forward
jet events. We note, however, that neutral higgsino production via VBF
has been suggested as a promising search channel after hard cuts on the
dijet invariant mass and $\eslt$ \cite{liantao}.} the projected event
rate is $< 1$~SS dilepton jj event/ab$^{-1}$ of integrated luminosity
even before acceptance, trigger and analysis cuts on $E_T(j)$ and jet
invariant masses that are necessary to extract the signal
\cite{ghww,aggies}.  Clearly the disparity between the cross sections
for higgsino-like and wino-like charginos cannot be explained by the
difference in the magnitudes of the $\th^0\th^\pm W^\mp$ and
$\tw^0\tw^\pm W^\mp$ couplings.  A resolution of this disparity forms
the subject of the remainder of this paper.

\begin{figure}[tbh]
{\begin{center}
\includegraphics[width=12cm,clip]{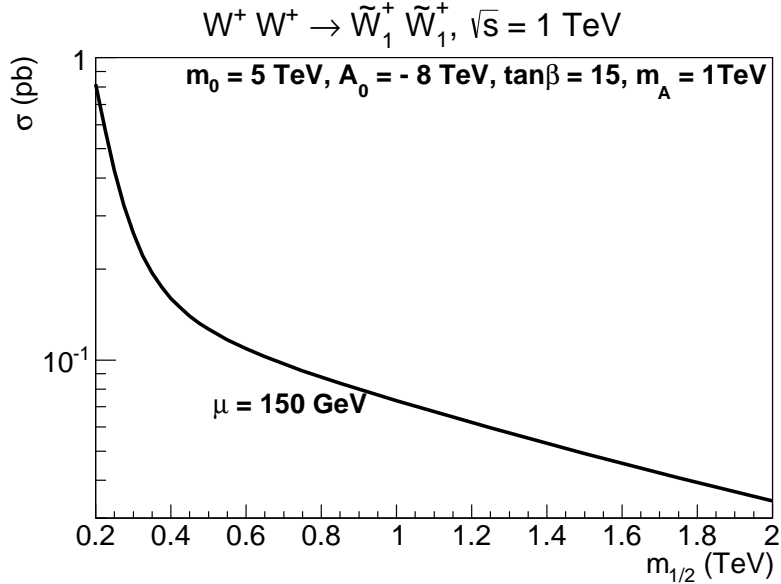}
\caption{The cross section for the underlying VBF process $W^+W^+\to
\tw_1^+\tw_1^+$ versus $m_{1/2}$ for the RNS
  model line (\ref{eq:mline}) 
of the text, assuming a centre-of-mass energy of 1~TeV. Notice that the
chargino mass is close to $\mu=150$~GeV over essentially the entire plot.}
\label{fig:WWtow1w1vsmhalf}
\end{center}}
\end{figure}

Toward this end, we show in Fig.~\ref{fig:WWtow1w1vsmhalf} the VBF cross
section for the underlying sub-process $W^+W^+ \to \tw_1^+\tw_1^+$ versus
$m_{1/2}$ for the same model-line as for the solid, black curve in
Fig.~\ref{fig:SSw1w1vsmhalf}, taking $\sqrt{s}=1$~TeV. In the MSSM,
this process occurs via the exchange of the four neutralinos in the $t$-
and $u$-channels. It is easy to see that the Majorana nature
of the neutralinos is essential to obtain a non-vanishing
amplitude. Indeed SS chargino production from VBF has been stressed as a
definitive test of Majorana nature of neutralinos \cite{tilman}.  As
before, we see that the cross section rapidly decreases with increasing
$m_{1/2}$, even though the produced chargino mass is close to 150~GeV
across the entire plot. The figure makes it obvious that the cross
section for same sign higgsino pair production is being {\em dynamically
suppressed}.

To better understand this suppression, we examine the interactions of
the chargino-neutralino system with $W$ bosons in the limit of large
gaugino masses, the situation we expect in natural SUSY. In this limit,
the winos and binos essentially decouple, leaving the charged (Dirac)
higgsino, $$\tw_1\equiv
(-i\gamma_5)^{\theta_\mu+1}\left(P_L\psi_{h_d^-}-P_R\psi_{h_u^+}\right),$$
and the two Majorana neutralinos,
$$(-i\gamma_5)^{\theta_\mu}\frac{(\psi_{h_u^0} +
  \psi_{h_d^0})}{\sqrt{2}} \
{\rm and} \  (-i\gamma_5)^{\theta_\mu
+1}\frac{(\psi_{h_u^0}-\psi_{h_d^0})}{\sqrt{2}}\;,$$ each with mass $|\mu|$
in the spectrum. Here,  $\psi_{h_u^0}$ and $\psi_{h_d^0}$ are the Majorana
higgsino fields in the notation of Ref.\cite{wss}, 
$\theta_\mu =0$, if $\mu > 0$ and $\theta_\mu=1$
if $\mu < 0$. One can then combine
the two degenerate neutralinos into a single Dirac higgsino-neutralino
$$\tz_D \equiv (-i\gamma_5)^{\theta_\mu+1}
\left(P_L\psi_{h_d^0}-P_R\psi_{h_u^0}\right)$$ with mass $|\mu|$.  
The masses of the charged and neutral Dirac higgsinos and their
couplings to the $W$-boson can readily be worked out. In the limit that
the gauginos are completely decoupled, we find, 
\be
{\cal L} = -|\mu| (\overline{\tw_1}{\tw_1}+\overline{\tz_D}\tz_D) +
\left[\frac{g}{\sqrt{2}}\overline{\tw_1}\gamma^\mu\tz_D
W_\mu+h.c.\right].
\label{eq:lag}
\ee
We see that the Lagrangian in Eq.~(\ref{eq:lag}) respects a new global
$U(1)_{\rm ino}$ symmetry, and so conserves ino-number, defined to be +1
for the Dirac particles $\tw_1$ and $\tz_D$, -1 for the corresponding
anti-particles, and 0 for sfermions and all SM particles. It is also
clear that it is not possible to consistently define $U(1)_{\rm ino}$
transformations unless the two neutral higgsino states are exactly
degenerate.  Ino-number conservation then requires the cross section for
the process $W^+W^+ \to\tw_1^+\tw_1^+$ must vanish in the limit
$M_{1,2}\to \infty$, accounting for the fall-off of the cross section in
Fig.~\ref{fig:WWtow1w1vsmhalf}. In the original language of two light,
neutral Majorana neutralinos, it is straightforward to analytically see
that the amplitude from the exchange of $\tz_1$ exactly cancels that
from the exchange of $\tz_2$ in the limit $M_{1,2} \to \infty$ in both
the $t$- and the $u$-channels: the magnitudes of the $\tz_i\tw_1^\pm
W^\mp$ ($i=1,2$) couplings and masses are identical for each of these
amplitudes, and the destructive interference arises because the two
neutralinos necessarily have opposite signs of the eigenvalue of the
neutralino mass matrix.

The couplings of the higgsinos to the fermion-sfermion system do not
respect the $U(1)_{\rm ino}$ invariance and so violate ino number
conservation. However, since first and second generation masses are only
weakly constrained by naturalness considerations, we expect that the
sfermion-mediated amplitudes in Fig.~\ref{fig:SSw1w1} will be strongly
suppressed (recall that we took first/second generation squark masses to
be 5~TeV in our illustrative examples) so that (approximate) ino-number
conservation once again accounts for the strong suppression of the SS chargino
production cross section for large values of electroweak gaugino mass
parameters at LHC13. Although obvious, we note that ino-number
conservation {\em does not} constrain $\tw_1^+ \tw_1^-$, $\tw_1 \tz_i$
and $\tz_i\tz_j$ production since the final states can have zero
ino-number.  

In summary, we have studied prospects for detecting the light higgsinos
that are expected to be the most robust characteristic of at least the
simplest models of natural SUSY at the LHC.  We focussed our attention
on the production of SS higgsinos produced via $pp \to
\tw_1^\pm\tw_1^\pm jj+X$ which, we anticipated would be dominated by VBF
processes because the $W$-boson couples to the higgsino system with the
{\em full-strength} $SU(2)$-doublet coupling. In natural SUSY models,
the signature would be a pair of same-sign, low $p_T$ leptons in events
with two forward high $E_T$ jets with a large rapidity separation
between them. We found that the cross section for the process is
typically smaller than 0.02~fb at LHC13 (two orders of magnitude smaller
than the corresponding cross section for SS wino production), even
before leptonic branching fractions of the charginos, or acceptance,
trigger and analysis cuts are folded in. We traced this suppression to
an approximate $U(1)_{\rm ino}$ symmetry that emerges in natural SUSY
models when $|\mu| \ll M_{1,2}$, or equivalently, to a (nearly) complete
cancellation between amplitudes from $\tz_1$ and $\tz_2$ exchange (the
amplitudes from other two neutralinos are suppressed because these are
heavy). We conclude that while chargino and neutralino production via
VBF processes may lead to observable signals at the LHC, it will not be
possible to search for the light charged higgsinos of natural SUSY even
at the high luminosity LHC via SS chargino production by VBF (unless
electroweak gaugino mass parameters are also fortuitously small -- in
which case there will be signals in several other channels) because the
signal is severely rate-limited.

\section*{Acknowledgments}
We thank H.~Baer and A.~Mustafayev for helpful comments on the text.
This work
was supported in part by the US Department of Energy.

%


\begin{thebibliography}{99}

\bibitem{atlas_susy} G.~Aad {\it et al.}  [ATLAS Collaboration], {\em J. High
Energy Phys.} {\bf 1409} (2014) 176 and
{\em J. High Energy Phys.} {\bf 1504} (2015) 116; 
G.~Aad {\it et al.} (ATLAS Collaboration)
\jhep{1405}{2014}{071} and \epjc{75}{2015}{208}.



\bibitem{cms_susy} S.~Chatrchyan {\it et al.}  [CMS Collaboration], {\em
  J. High Energy Phys.} {\bf 1406} (2014) 055; V.~Khachatryan {\it et
  al.} [CMS Collaboration] {\em J. High Energy Phys.} {\bf 1505} (2015)
  078; V.~Khachatryan {\it et al.} (CMS Collaboration)
  \epjc{74}{2014}{3036} and {\em Phys. Rev.} {\bf D90} (2014) 092007;


%
\bibitem{wessz} Y.~Golfand and E.~Likhtman, {\em JETP Lett.} {\bf 13}
  (1971) 323; D.~Volkov and V.~Akulov, {\em JETP Lett.} {\bf 16}
  (1972) 621; J.~Wess and B.~Zumino, {\em Nucl. Phys.} {\bf B70} (1974)
  39. 


%
\bibitem{hier} E.~Witten, {\em Nucl. Phys.} {\bf B188} (1981) 513;
S.~Dimopoulos and H.~Georgi, {\em Nucl. Phys.} {\bf B193} (1981),150;
N.~Sakai, {\em Z. Phys.} {\bf C11} (1981) 153;  R.~Kaul, {\em Phys. Lett.} {\bf B109} (1982) 19.
%

\bibitem{susswein} E.~Gildener and S.~Weinberg,  {\em Phys. Rev.} {\bf D13} 
(1976) 3333; E.~Gildener, {\em Phys. Rev.} {\bf D14} 
(1976) 1667; L.~Susskind, {\em Phys. Rev.} {\bf D20} (1979) 2619.

%
\bibitem{ftune} For early discussions, see {\it e.g.}
   R.~Kitano and Y.~Nomura, {\em Phys. Lett.} {\bf B631} (2005) 58  and 
  {\em Phys. Rev.} {\bf D73} (2006) 095004; For more recent studies in
   light of LHC results, see
M.~Papucci, J.~T.~Ruderman and A.~Weiler,
{\em J. High Energy Phys.} {\bf 1209} (2012) 035;
A.~Strumia, {\em J. High Energy Phys.} {\bf
  1104} (2011) 073; J.~Lykken and M.~Spiropulu,
 {\it  Sci. Am.} {\bf 310N5} (2014) 5,  36;  N.~Craig,
  arXiv:1309.0528 [hep-ph]. 

%
\bibitem{lhcstop} CMS Collaboration, S.~Chatrchyan {\it et. al.}
  \epjc{73}{2013}{2677} and V.~Khachatryan {\it et al.} arXiv:1503.08037
  [hep-ex]; ATLAS Collaboration, G.~Aad {\it et al.} arXiv:1506.08616
  [hep-ex]. 

%
\bibitem{rnsltr}  H.~Baer, V.~Barger, P.~Huang, A.~Mustafayev and X.~Tata,
 {\em Phys. Rev. Lett.} {\bf 109} (2012) 161802.

%
\bibitem{rns} H.~Baer, V.~Barger, P.~Huang, D.~Mickelson, A.~Mustafayev
  and X.~Tata, {\em Phys. Rev.} {\bf D87} (2013) 115028.

%
\bibitem{howieover}H.~Baer, V.~Barger and D.~Mickelson, {\em Phys. Rev.}
{\bf D88} (2013) 095013; H.~Baer, V.~Barger and M.~Savoy, {\em Physica
Scripta} {\bf 90} (2015) 6, 068003, arXiv:1502.04127.

%
\bibitem{am} A.~Mustafayev and X.~Tata, invited contribution in volume
  commemorating C.V. Raman's 125th birth anniversary, {\em Indian
  J. Phys.} {\bf 88} (2014) 991; X.~Tata, arXiv:1506.07151 [hep-ph].
%
\bibitem{CCN} K.~Chan, U.~Chattopadhyay and P.~Nath, {\em Phys. Rev.}
{\bf D58} (1998) 096004.
%
\bibitem{nelmar}A.~Nelson and T. Roy, {\em Phys. Rev. Lett.} {\bf 114}
(2015) 201802; S.~Martin, arXiv:1506.02105 [hep-ph]. 
%
\bibitem{luty} T.~Cohen, J.~Kearney and M.~Luty, {\em Phys. Rev.} {\bf
  D91} (2015) 075004.

%
\bibitem{rnslhc} H.~Baer, V.~Barger, P.~Huang, D.~Mickelson,
  A.~Mustafayev, W.~Sreethawong and X.~Tata {\em J. High Energy Phys.} {\bf
  1312} (2013) 013 and {\em  J. High Energy Phys.} {\bf
  1506} (2015) 053 (Erratum).
%
\bibitem{ilc} H.~Baer, V.~Barger, D.~Mickelson, A.~Mustafayev and
  X.~Tata, \jhep{1406}{2014}{172}.
%
%
\bibitem{nugm} H.~Baer, V.~Barger, P.~Huang, D.~Mickelson,
  M.~Padeffke-Kirkland and X.~Tata, {\em Phys. Rev.} {\bf D91} (2015)
  075005.
%
\bibitem{mono} H.~Baer, A.~Mustafayev and X.~Tata, \prd{89}{2014}{055007};
C.~Han, A.~Kobakhidze, N.~Liu, A.~Saavedra, L.~Wu  and J.~M. Yang,
 {\em J. High
Energy Phys.} {\bf 1402} (2014) 049.
P.~Schwaller and J.~Zurita,  {\em J. High
Energy Phys.} {\bf 1403} (2014) 060; D.~Barducci, A.~Belyaev,
A.~Bharucha, W.~Porod and V.~Sanz, arXiv:1504.02472 express a more
optimistic viewpoint for detection of the monojet signal.
%
\bibitem{carp} A.~Anandakrishnan, L.~Carpenter and S.~Raby,
  \prd{90}{2014}{055004}. 
%
%
\bibitem{ghww} G~Giudice, T.~Han, K.~Wang and L-T.~Wang, {\em
  Phys. Rev.} {\bf D81} (2010) 115011.
%
\bibitem{kribs}Z.~Han, G.~Kribs, A.~Martin and A.~Menon,  {\em
  Phys. Rev.} {\bf D89} (2014) 075007.
%
\bibitem{dilep} H.~Baer, A.~Mustafayev and X.~Tata,  {\em
  Phys. Rev.} {\bf D90} (2014) 115007.
%
\bibitem{atlasWW} G.~Aad {\it et al.} (ATLAS Collaboration)
{\em Phys. Rev. Lett.} {\bf 113} (2014) 141803.
\bibitem{cmsWW} V.~Khachatryan {\it et al.} (CMS Collaboration)
{\em Phys. Rev. Lett.} {\bf 114} (2015) 051801.


%
\bibitem{tilman} G.~Cho, K.~Hagiwara, J.~Kanzaki, T.~Plehn, D.~Rainwater and T.~Stelzer,  {\em
  Phys. Rev.} {\bf D73} (2006) 054002.
%
\bibitem{aggies} A.~Delannoy {\it et al.} {\it Phys. Rev. Lett.} {\bf
  111} (2013) 061801; B.~Dutta {\it et al.} {\em
  Phys. Rev.} {\bf D87} (2013) 035029; B.~Dutta {\it et al.} {\em
  Phys. Rev.} {\bf D90} (2014) 095022; B.~Dutta {\it et al.} {\em
  Phys. Rev.} {\bf D91} (2015) 055025.
%
%
\bibitem{nuhm2} D.~Matalliotakis and H. P.~Nilles, {\em Nucl. Phys.} {\bf
    B435} (1995) 115; V.~Berezinsky, A.~Bottino, J.~Ellis, A.~Fornengo,
    G.~Mignola and S.~Scopel, {\em Astropart. Phys.}
    {\bf 5} (1996) 1; P.~Nath and R.~Arnowitt, {\em Phys. Rev.} {\bf D56}
    (1997) 2820; J.~Ellis, K.~Olive and Y.~Santoso,
    {\em Phys. Lett.} {B539} (2002) 107; 
J.~Ellis, T.~Falk, K.~Olive and Y.~Santoso,
    {\em Nucl. Phys.} {\bf B652} (2003) 259; 
  H.~Baer, A.~Mustafayev, S.~Profumo, A.~Belyaev
    and X.~Tata, {\em J. High Energy. Phys.} {\bf 0507} (2005) 065.

%
\bibitem{isajet} ISAJET, H.~Baer, F.~Paige, S.~Protopopescu and X.~Tata,
  hep-ph/0312045. 
%
\bibitem{madgraph} Madgraph 5, J.~Alwall, M.~Herquet, F.~Maltoni,
  O.~Mattelaer and T.~Stelzer, \jhep{1106}{2011}{128};
  J.~Alwall {\it et al.}, \jhep{1407}{2014}{079}.

%
\bibitem{pythia} PYTHIA 6.4, T.~Sjostrand, S.~Mrenna and P.~Skands,
  \jhep{0605}{2006}{026} [hep-ph/0603175].
%
\bibitem{cteq} J.~Pumplin, D.~Stump, J.~Huston, H.~Lai, P.~Nadolksy and
W.-K.~Tung, \jhep{0207}{2002}{012}. 
%
\bibitem{pgs} PGS, J.~Conway, http://www.physics.ucdavies.edu/~conway/research/software/pgs/pgs4-general.htm.
%
\bibitem{sppc} M.~Ahmad {\it et al.} (CEPC-SPPC Study Group) 
http://cepc.ihep.ac.cn/preCDR/volume.html
%
\bibitem{liantao} A.~Berlin, T.~Lin, M.~Low and L-T.~Wang,
  \prd{91}{2015}{115002}. 
%
\bibitem{wss} H.~Baer and X.~Tata, {\em Weak Scale Supersymmetry}
  (Cambridge, 2006).
\end{thebibliography}
\end{document}